# Utility-Scale Bifacial Solar Photovoltaic System: Optimum Sizing and Techno-Economic Evaluation

Sharaf K. Magableh, Caisheng Wang, Feng Lin
Department of Electrical and Computer Engineering
Wayne State University
Detroit, United States
sharaf.magableh@wayne.edu, cwang@wayne.edu, flin@wayne.edu

*Abstract*—Classical monofacial solar photovoltaic systems have gained prevalence and are widely reported in the literature because they have a lower initial cost compared with bifacial systems. However, limited investigation of both systems has been done on a utility scale with different performance indicators. This paper introduces a multifaceted comparative analysis including various aspects like energy generation, reliability, environmental effect, economic viability, and footprint area. Real measured data, including ambient temperature, solar irradiance, and a utility-scale load, were used for studying both systems in the City of Detroit. The optimal system sizing and energy management strategy are attained using the Whale optimization algorithm. Minimizing the loss of power supply probability and sizing the number of photovoltaic panels ($N_{PV}$) are carried out for both cases. Results revealed that the bifacial solar system generates more power with a lower $N_{PV}$, a smaller installation area, and hence a lower levelized cost of energy for the entire project lifetime compared to the monofacial system. Accordingly, the bifacial system outlined in this paper is recommended and can be implemented in various locations to establish a sustainable solar energy system that is economically feasible with clean energy production for the entire project's lifespan.

## I. Introduction

In the past decade, photovoltaic (PV) production has experienced extreme growth globally, as the cumulative installed capacity of PV reached 1.117 TW in 2022 [1]. This fast evolution is because PV is a green, low-maintenance, inexhaustible, silent, and publicly acceptable energy generation technology [2]. However, among several PV categories, the conventional monofacial PV (mPV) technology is not capable of absorbing all the obtainable solar irradiance to generate electricity, which eventually lowers the efficiency [3]. On the other hand, bifacial PV (bPV) panels absorb solar irradiance from both front and rear sides, hence enhancing the efficiency of the entire PV system. The reflected radiation from Earth's albedo on the rear side of the PV panel could increase the generated electricity. Therefore, the trend of Crystalline Silicon cells in the industry and academic research has been shifted from mPV to bPV. The global market share of bPV was around 30% in 2022, and it is predicted to increase to 70% in 2030 [4].

In the existing literature, several research efforts studied the application of bPV technology. In [4], the authors explored the challenges associated with the bPV technology compared with the conventional mPV systems. They studied issues related to irregular backside irradiance and other limitations that bulk the effectiveness of bPV. Accordingly, they proposed potential future approaches to enhancing the productivity of bPV technology. Their results revealed that utilizing certain bifacial construction for bPV modules could yield an increase of 5% to 30% in the output power with an increase of 15.6% in the initial cost, and would result in a net reduction of the levelized cost of energy by 2% to 6% compared to mPV systems. The authors concluded that for optimum electricity generation, bPV modules should be installed within areas of high albedo and elevation. This is because such configurations enable these modules to exploit light from both sides, thus generating extra power. Nonetheless, their proposed system has not been optimized within an actual grid-connected renewable energy framework, considering different performance assessments. Researchers of [5] analyzed the sensitivity of bPV and mPV solar panels to determine the optimum performance conditions for the bPV panels. This was done using the PVsyst software at different sensitivity parameters such as different tilt angles of (15°, 25°, 35°), albedos of (10%, 50%, 90%), and heights of (0.5m, 1m, 1.5m) above the Earth's surface. They found that the generated energy increases when the values of albedo and height above the Earth's surface are higher. The best tilt angle in the case of bPV was 35° at high values of albedo and the highest installed point above the ground compared with 25° tilt angle in the case of mPV. The average daily bifacial gain of 36.68% occurred at 35° tilt angle, albedo of 90%, and 1.5m height. Although they studied the effect of various weather parameters on the sensitivity of bPV and mPV technologies, the impacts of a utility-level bPV or even mPV systems have not been fully taken into consideration, and the optimization in terms of sizing, reliability, area, or even ecological impacts have not been thoroughly studied. Therefore, this paper compares the effectiveness of installing bPV and mPV at the utility scale including several performance indicators. In [6], the authors developed a sophisticated incident model for a fixed multi-row bPV systems and accounting various paraments including ground albedo. Their model incorporates the anisotropic sky model, and sun angles under several conditions. Their analytical model is able to compute the solar irradiance on the front and rear sides every time, everywhere. They simulate their system on the IEEE 69-bus system to express how bPV can mitigate the "Duck curve" problem produced by conventional mPV systems. The findings reveal that a well-aligned and elevated bPV system can produce up to 31% more yearly energy than a mPV system. In [7], the researchers examined the improvement in the output power performance of a bPV system using a new model by applying various surfaces to test the extracted output power. This model considers the solar radiation contributions, including ground-reflected irradiance from diffused horizontal irradiance (DHI) and direct normal irradiance (DNI), and their effects on both front and back sides. The finding showed that the annual energy production decreased by 4% when a shadow existed compared with the case of no shadow. They also found that the bifacial gain was reduced to less than 6% while simulating a dry asphalt surface (albedo of 12%) compared with a reduction of 29% using a white-covered surface (albedo of 70%). Although they considered different cases of bPV, they did not compare with other solar technologies, i.e., mPV, when computing the extracted energy. So, in this paper, the effect of solar irradiance on mPV and bPV solar technologies is investigated. This is in terms of the system's sizing and technical performance evaluation on a utility-scale power system.

This paper makes contributions in several aspects as follows. This study provides an inclusive comparison of grid-connected solar renewable energy systems, specifically exploring the implementation of mPV and bPV solar modeling on a Megawatt scale in Detroit City, Michigan. The focus was on a detailed modeling of the PV power output for both solar systems, using real measured data to optimize each case. Through detailed technical assessments, the best-case scenario, determining design configurations, and evaluating output energies have been investigated, aiming to compare optimal sizing in terms of reliability, cleaner production, geographical installation area, economic feasibility, and various technical indicators. The grid-connected PV systems, including mPV and bPV solar technologies, are investigated as illustrated in Fig. 1.

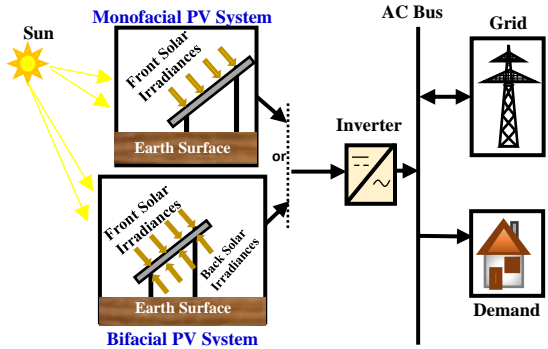

Fig. 1. The on-grid solar PV system applying either mPV or bPV technologies.

## II. SYSTEM'S DATA

The real measured data are vital to get the practical and precise design of the solar system in both solar mPV and bPV arrays. The global horizontal irradiance (GHI), DNI, and DHI in ($W/m^2$), and solar ambient temperature ($T_{amb}$) in (℃) were obtained from the National Solar Radiation Database (NSRDB). These data are acquired for Wayne State University (WSU) campus, Detroit, USA, to simulate the proposed system in Fig. 1. Fig. 2 shows that GHI, DNI, and DHI have a maximum value of around 1028, 1024, and 474 ($W/m^2$), respectively, and $T_{amb}$ has a maximum value of 31.8 (℃). This indicates that WSU territory has the potential to exploit these solar energy resources, which will achieve promising results regarding reliability and ecological impact.

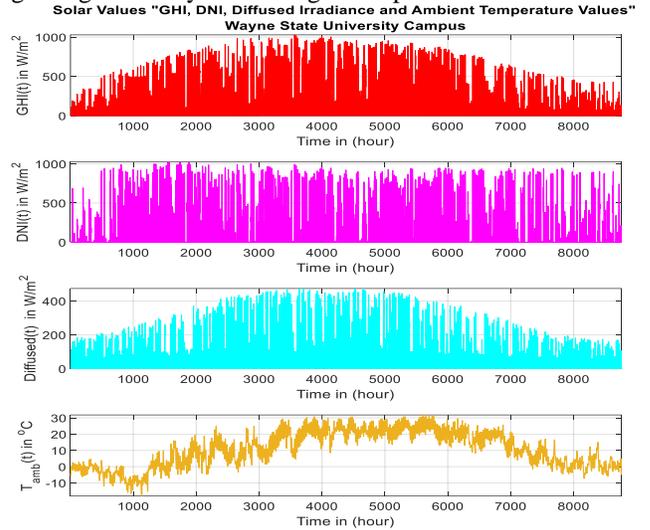

Fig. 2. Hourly measured solar data at WSU Campus, Michigan in 2021.

Subsequently, the hourly measured load for the 22kV sub-feeder near the WSU campus in Detroit is used to design the mPV and bPV solar grid-connected systems, shown in Fig. 3. Noted that the maximum value of the load demand is 1.7975MW and the average value is 1.0096MW.

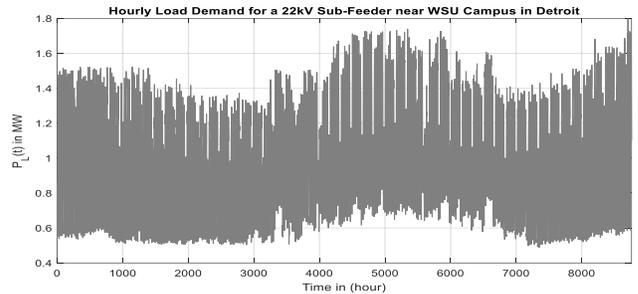

Fig. 3. Hourly measured 22kV sub-feeder load in Detroit in 2021.

## III. MATHEMATICAL MODELING AND DESIGN METHODOLOGY

Mathematical modeling for each component of the proposed system is a crucial step in the design of the overall system. Therefore, the hourly horizontal solar-measured real data at the WSU Campus presented in section II will be employed in the modeling of the PV panel angles. It should be noted that the modeling of solar angles, mPV, and bPV are explained in detail in [8]. This is to find the hourly global tilted solar irradiance for both mPV ($I_{Gtm}(t)$) and bPV ($I_{G_B}(t)$).

### A. mPV and bPV Output Irradiances for a Panel

In this paper, the bPV module is chosen as a new PV technology to compare its performance with the traditional mPV technology. The rated power of the chosen bifacial PV module is 462W. Hence, based on the input solar data from the manufacturer datasheet, and the selected solar PV panel data sheet [8], global tilted irradiance for mPV and bPV ($I_{Gtm}$) and ($I_{G_B}$) are computed accordingly. The bPV solar panels absorb the irradiance from both front and rear sides, by executing the global beam, diffused, and reflected irradiances ($I_{G_{B_{Beam}}}$),

($I_{G_{B_{Diffused}}}$), and ($I_{G_{B_{Reflected}}}$) data and eventually compute the tilted $I_{G_B}$. Hence, the tilted front solar irradiances are obtained as illustrated in Fig. 4.

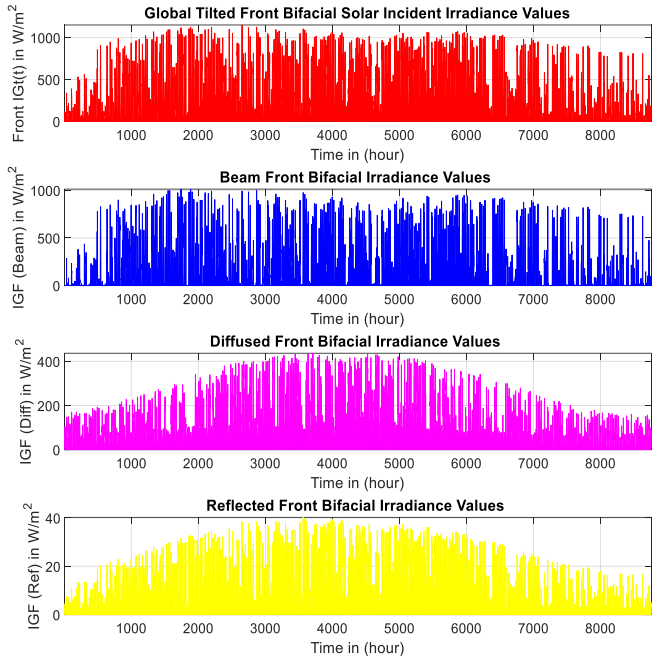

Fig. 4. Hourly tilted front bifacial solar irradiance values in W/m².

The values of the back irradiances are lower than the front irradiances, as shown in Fig. 5. Note that all these variables are calculated at each time step depending on the sun's variable altitude at the specific location for the entire 8760 hours of the year 2021.

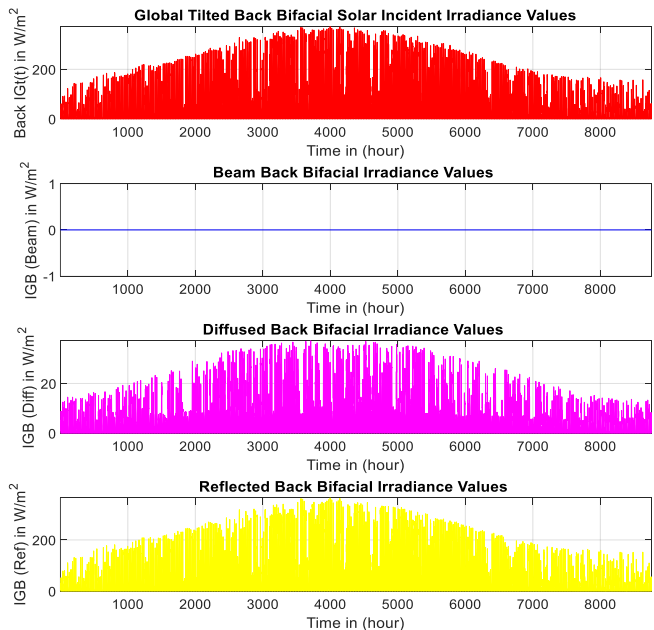

Fig. 5. Hourly tilted back-side bifacial solar irradiance values in W/m².

It can be seen that the values of the front global beam irradiance ($I_{G_{B_{Beam}}}$) are higher compared with the front $I_{G_{B_{Diffused}}}$, and front $I_{G_{B_{Reflected}}}$. In addition to the fact that any solar irradiance values of beam, diffused, and reflected are lower in winter than in summer. The front $I_{G_B}$ reached its peak of around $1,150 W/m^2$ compared with $372.2 W/m^2$ for the rear side. It can also be noticed that the direct rear $I_{G_{B_{Beam}}}$ is always zero.

### B. mPV and bPV Global Tilted Solar Irradiances

Subsequently, the global tilted irradiance for bPV exceeds that of mPV. This is accredited to bPV's ability to absorb irradiance from both its front and back sides, as depicted in Fig. 6. The maximum value of $I_{Gtm}$ reaches $1143.6 W/m^2$, compared with $I_{G_B}$ of $1380.46 W/m^2$, as shown in Fig. 6. Moreover, the mean values of $I_{G_{tm}}$ and $I_{G_{tb}}$ are $205.84 W/m^2$ and $234.42 W/m^2$, respectively. When comparing bPV to mPV, there is a corresponding percentage increase of 20.71% for maximum tilted solar irradiance and around 13.88% for the mean values.

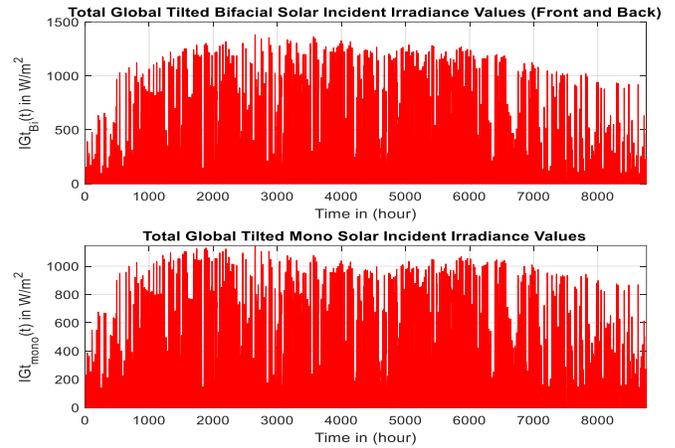

Fig. 6. Global hourly tilted monofacial and bifacial irradiance values in W/m².

Certainly, the percentage difference will have several merits for implementing bifacial technology in practical solar applications. This will result in a notable reduction in the number of PV panels in the entire solar system, especially for large-scale MW projects, consequently reducing the required geographical installation area.

## IV. ENERGY MANAGEMENT APPROACH, SYSTEM'S OPERATIONAL SCHEME AND PERFORMANCE INDICATORS

The system's operational scheme outlines the power generation strategy of the PV system for each time interval, as shown in Fig. 7.

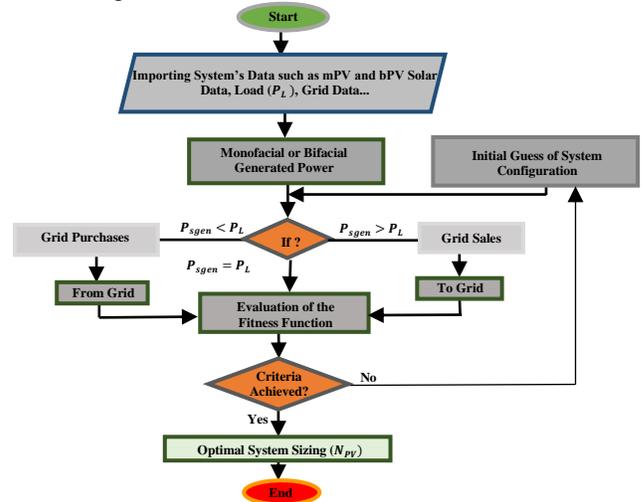

Fig. 7. System's energy management and operational strategy.

The DC electricity generated from the PV panels needs to be converted into AC ($P_{sgen}(t)$), as illustrated in (1). It is important to note that if the generated power from the PV system is less than the load demand ($P_L(t)$), any deficit will be covered by the utility grid, known as the purchased power ($P_{gpurch}(t)$) computed as in (2). Conversely, when there is extra power, or the PV generation is greater than the load, the surplus power is sold back to the utility ($P_{gsold}(t)$) as in (3). The complete balance equation used to accurately size the solar PV system at each hour is shown in (4).

$$P_{sgen}(t) = \eta_{inverter} \times P_{PV}(t) \times f_{PV} \quad (1)$$
$$P_{gpurch}(t) = P_L(t) - P_{sgen}(t) \quad (2)$$
$$P_{gsold}(t) = P_{sgen}(t) - P_L(t) \quad (3)$$
$$P_{sgen}(t) + P_{gpurch} = P_L(t) + P_{gsold}(t) + P_{deficit}(t) \quad (4)$$

### A. Objective Function

In this paper, the objective is to minimize the loss of power supply probability (LPSP) using the whale optimization algorithm (WOA). The WOA is chosen because it has a good convergence behavior compared with other state-of-the-art algorithms [9]. The optimization process is conducted on the proposed methodology to simulate the proposed system in Fig. (7). LPSP is a metric that measures how much the energy generated by the system fails to meet the load, as specified in equation (5) [10]. The lowest LPSP value indicates more efficient power generation from the PV system. The minimization process of LPSP and sizing of the decision variable, the number of solar PV panel's ($N_{PV}$), is subjected to constraints outlined in (6). The optimized objective function is subjected to inequality and quality constraints as shown previously in (2), (3), and (4), respectively. Note that, $P_{gpurch_{max}}$ is the maximum allowable purchased power from the grid at each time step.

$$LPSP = \frac{\sum_{t=1}^{8760}(P_L(t) - (P_{sgen}(t) + P_{gpurch}(t)))}{\sum_{t=1}^{T} P_L(t)} \quad (5)$$

$$\begin{cases} \underset{(N_{PV})}{\text{Minimization}} \langle LPSP \rangle = Min\left(\frac{\sum_{t=1}^{8760} P_L(t) - (P_{sgen}(t) + P_{gpurch}(t))}{\sum_{t=1}^{T} P_L(t)}\right) \\ Subjec\ to: \\ \begin{cases} P_{gpurch} \leq P_{gpurch_{max}} \\ Equations\ (2),(3), and\ (4) \end{cases} \end{cases} \quad (6)$$

### B. Additional Performance Indicators

This section addresses a set of performance indicators that are employed to evaluate the optimal sizing of both mPV and bPV systems in terms of reliability, economic viability, and environmental impact. These metrics are the carbon dioxide ($CO_2$) reduction amount ($CO_2RA$), and the levelized cost of energy (LCOE). $CO_2RA$ quantifies the reduction in the harmful gas emissions achieved by implementing solar PV energy, as in (7) [11]. A greater $CO_2RA$ value reflects a more effective utilization of solar renewable resources and, hence larger reduction of system's emissions. Note that, $F_{CO_2}$ in (7) represents the carbon dioxide emission factor, which is around 0.553 tCO2/MWh in Michigan [12].

$$CO_2RA = E_{sgen} \times F_{CO_2} \quad (7)$$

This study evaluates the economic feasibility of both the mPV and bPV energy systems by computing the LCOE. The goal is to determine the energy price needed for the mPV and the bPV systems to generate and meet the electricity demand of the mentioned load in section II. The LCOE is computed by dividing the total annualized cost (TAC) by the energy produced to meet the load demand ($E_g$), as outlined in equation (8). The value of $E_g$ is determined at the optimal LPSP value, which was optimized using the WOA. The cost values of each component in the PV system, including the cost parameters that are required to build the nominal and discounted cashflows with 25 years lifetime, are illustrated in detail in [8, 13].

$$LCOE = \frac{TAC}{E_g} \quad (8)$$

### C. Needed PV Installation Footprint Area

The calculation of the necessary area for the installation of the PV array is computed in (9) [14]. This calculation takes into account several factors, including the PV module ($A_m$), the number of columns ($N_{col}$) which is equivalent to $N_{PV}$ connected in series per string as in (10), and tilt angle. Note that, $N_{rows}$ is the number of rows or strings of a PV array [15].

$$A_{PVplant} = A_m N_{PV} \cos(\beta) + 3A_m(N_{PV} - N_{col})\sin(\beta) \quad (9)$$
$$N_{col} = \frac{N_{PV}}{N_{rows}} \quad (10)$$

## V. RESULTS AND DISCUSSION

This section implements the mathematical modeling and the real measured data from the WSU Campus area to establish the optimum configuration for the mPV and the bPV systems. The approach involves employing the WOA to minimize LPSP and, as a result, determining the optimal system sizes for both scenarios. Table I illustrates the optimal dimensions for the mPV and the bPV solar grid-connected systems. Table I presents that the number of bPV panels is lower than that of mPV panels. This means that even if bPV panels are more expensive, they require a lower number of installed panels, which will affect the entire cost and footprint area of the system. The reason for this is that the solar bPV technology generates a greater amount of power compared to mPV, and this means it needs fewer solar units. The LPSP values are around 0.634% and 0.5757% for the mPV and the bPV technologies, respectively. This indicates that the bPV system operates more effectively to meet the demand. Accordingly, the WOA iterative curves of the optimum objective LPSP for both cases are shown in Fig. 8.

TABLE I. OPTIMUM VALUES OF THE OBJECTIVE FUNCTION AND THE DECISION VARIABLE FOR BOTH CASES

| Cases | $N_{PV}$ | LPSP in % |
|---|---|---|
| mPV | 12,253 | 0.634 |
| bPV | 10,924 | 0.5757 |

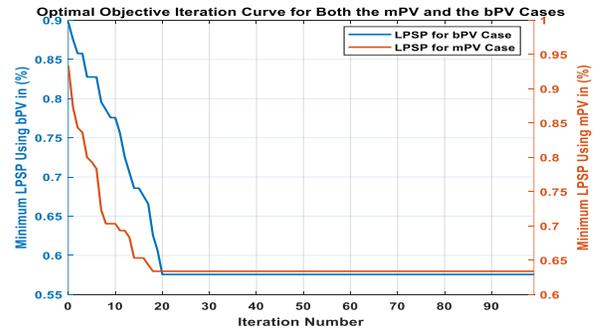

Fig. 8. The optimum LPSP for both the mPV and the bPV systems.

Table II illustrates ecological and economic performance indicators in addition to the required installation areas for both mPV and bPV cases. Note that these values were computed at the optimum case of LPSP and the optimal size. Table II shows that the bPV has reduced $CO_2$ emissions by 6.2896GgCO$_2$/year.

This is a difference of 13.88% from the mPV technology, which proves that the bPV is more environmentally friendly by reducing the harmful emissions produced by the utility and replacing it with cleaner energy. Subsequently, the LCOE is around 0.24695$/kWh and 0.27213$/kWh for the bPV and the mPV systems, respectively. This means that the bPV is more economically feasible than the mPV system. Although bPV panels have higher capital cost, it has lower LCOE for the entire system cost. This is because it has a lower $N_{PV}$, generates more electricity, and reduces operation and maintenance costs. Moreover, the required solar PV plant area to construct the mPV and the bPV arrays are around 16.31 acres and 14.54 acres, respectively. Hence, the bPV system needs a smaller installation area than the mPV plant because it absorbs irradiance from both sides and generates more energy. This indicates the feasibility of adopting the bPV technology and proves how it has a lower LCOE during the lifespan of the project and footprint area compared with the classical mPV.

TABLE II. SYSTEM'S ECOLOGICAL IMPACT, AREA, AND COST ANALYSIS VALUES AT THE OPTIMAL CONFIGURATION FOR BOTH CASES

| Indicator | Value | |
|---|---|---|
| | mPV | bPV |
| $CO_2RA$ (in Gg$CO_2$/year) | 5.4734 | 6.2896 |
| LCOE ($/kWh) | 0.27213 | 0.24695 |
| $A_{PV\,Plant}$ (Acre) | 16.31 | 14.54 |

Accordingly, Table III shows all energy values for both cases. In the bPV technology, it can be noticed that the purchased energy is 17.61GWh, and the sold energy is 1.632GWh. This indicates more solar PV energy utilization and more efficient performance to satisfy the load in the case of the bPV system compared with the mPV system.

TABLE III. ALL ENERGY VALUES IN (GWH) AT THE OPTIMAL CONFIGURATION FOR BOTH CASES

| Energies (GWh) | mPV | bPV |
|---|---|---|
| $E_{sgen}$ | 9.897 | 11.374 |
| $E_{gpurch}$ | 18.177 | 17.61 |
| $E_L$ | 27.511 | 27.511 |
| $E_{gsold}$ | 0.723 | 1.632 |
| $E_{deficit}$ | 0.15884 | 0.1587 |

The data in Fig. 9 displays the inverted solar PV output power values for the bPV and the mPV systems in MW. It can be noticed that bPV has the highest inverted power value of around 5.71MW compared with 4.817MW for the mPV system.

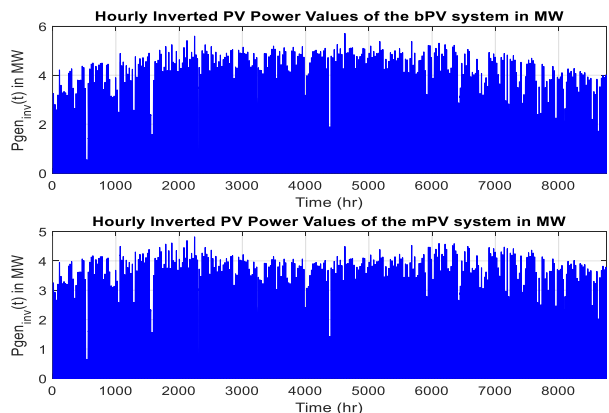

Fig. 9. Hourly inverted bPV and mPV power values.

Moreover, the average generated powers for the entire year were approximately 1.298MW and 1.129MW for the bPV and the mPV technologies, respectively. This indicates the ability of bPV systems to absorb solar irradiance from both sides, and generate more energy with a lower $N_{PV}$, resulting in a lower LCOE for the entire project lifetime.

VI. CONCLUSIONS

This paper compared utility-scale mPV and bPV systems and investigated the optimal sizing and configuration of such systems. The systems were optimized to minimize the LPSP using WOA for both cases. Real measured data were obtained and used to determine the best configurations for both systems. Several performance indicators indicate that bPV is economically affordable with a smaller footprint and more ecological benefits. The results revealed that, although bPV systems could have a higher initial capital cost compared with mPV systems, it has a lower LCOE during the project's lifetime. The methodology presented in this paper can be applied effectively in different locations when the weather and environmental information is available at those locations.

REFERENCES

[1] S. Enkhardt. BloombergNEF says global solar will cross 200 GW mark for first time this year, expects lower panel prices. *PV Magazine*, 2022
[2] S. Slade. Wind & Solar Energy Installation Data. Available: http://www.fi-powerweb.com/About.html (accessed on 4 July 2022).
[3] U. J. R. E. Stritih, "Increasing the efficiency of PV panel with the use of PCM," *Renewable Energy*, vol. 97, pp. 671-679, 2016.
[4] W. Gu, T. Ma, S. Ahmed, Y. Zhang, J. Peng,"A comprehensive review and outlook of bifacial photovoltaic (bPV) technology," *Energy Conversion and Management*, vol. 223, p. 113283, 2020.
[5] C. Ghenai, F. F. Ahmad, O. Rejeb, and A. K. J. S. E. Hamid, "Sensitivity analysis of design parameters and power gain correlations of bi-facial solar PV system using response surface methodology," *Solar Energy*, vol. 223, pp. 44-53, 2021.
[6] Mahdi Rouholamini, Lei Chen, and Caisheng Wang, "Incident Energy Based Model to Optimally Configure Bifacial PV Arrays," *IEEE Transactions on Sustainable Energy*, pp. 1242 – 1255, vol. 12, no. 2, April 2021.
[7] D. Chudinzow, J. Haas, G. Díaz-Ferrán, S. Moreno-Leiva, and L. J. S. E. Eltrop, "Simulating the energy yield of a bifacial photovoltaic power plant," *Solar Energy*, vol. 183, pp. 812-822, 2019.
[8] H. M. K. Al-Masri, O. M. Dawaghreh, and S. K. Magableh, "Realistic performance evaluation and optimal energy management of a large-scale bifacial photovoltaic system," *Energy Conversion and Management*, vol. 286, p. 117057, 2023/06/15/ 2023.
[9] S. Mirjalili and A. Lewis, "The whale optimization algorithm," *Advances in engineering software*, vol. 95, pp. 51-67, 2016.
[10] H. M. Al-Masri, S. K. Magableh, A. Abuelrub, "Output power computation and sizing of a photovoltaic array by advanced modeling," *Sustainable Energy Technologies and Assessments*, vol. 47, p. 101519, 2021.
[11] A. D. BANK, "Guidelines for estimating greenhouse gas emissions of Asian Development Bank projects: Additional guidance for clean energy projects,"*Development Bank: Mandaluyong, Philippines*, 2017, Available: http://dx.doi.org/10.22617/TIM178659-2.
[12] EPA greenhouse gas reporting and emission rates. Available: https://shorturl.at/hFU28 (accessed on 7 Sep 2023).
[13] S. Singh and S. C. Kaushik, "Optimal sizing of grid integrated hybrid PV-biomass energy system using artificial bee colony algorithm," *IET Renewable Power Generation*, vol. 10, no. 5, pp. 642-650, 2016.
[14] S. Chakraborty, P. K. Sadhu, and N. Pal, "Technical mapping of solar PV for ISM-an approach toward green campus," *Energy Science Engineering*, vol. 3, no. 3, pp. 196-206, 2015.
[15] H. M. Al-Masri, O. M. Dawaghreh, and S. K. Magableh, "Optimal configuration of a large scale on-grid renewable energy systems with different design strategies," *Journal of Cleaner Production*, p. 137572, 2023.